\documentclass[%
 reprint,
 amsmath,amssymb,
 aps,
]{revtex4-2}
\usepackage{tikz}
\usetikzlibrary{decorations.pathmorphing}
\usepackage{graphicx}
\usepackage{dcolumn}
\usepackage{bm}

\usepackage{color}
\usepackage{colortbl}

\begin{document}

\preprint{APS/123-QED}

\title{The diffusion equation for non-Markovian Gaussian stochastic processes}

\author{Alessandro Taloni}
 \email{alessandro.taloni@cnr.it}
 \affiliation{CNR--Consiglio Nazionale delle Ricerche, ISC, Via dei Taurini 19, 00185 Rome, Italy}
\author{Gianni Pagnini}%
\affiliation{BCAM--Basque Center for Applied Mathematics \& Ikerbasque--Basque Foundation of Science, 
Alameda de Mazarredo 14, 48009 Bilbao, Basque Country -- Spain
}
\author{Aleksei Chechkin}%
 \affiliation{Akhiezer Institute for Theoretical Physics, NSC KIPT, Kharkov 61108, Ukraine}

\date{\today}

\begin{abstract}
We derive the exact evolution equation for the probability density function (pdf) of particle displacements generated by general 
Gaussian velocity processes, 
when neither Markovianity and nor stationarity are assumed. 
Starting from the characteristic function of the displacements pdf, 
we construct a systematic hierarchy of equations based on Wick’s theorem, 
in which the dynamics is governed by sums of geometrically connected Wick contractions. 
This approach yields a closed non‑Markovian diffusion equation that generalizes 
the Fokker--Planck description and preserves Gaussianity only in the infinite‑order limit. 
\end{abstract}

\maketitle

The dynamics of a particle undergoing a stochastic motion with
velocity $v(t)$ is described by the governing equation of
its probability density function $P(v,t)$,
namely, by the Fokker-Planck equation 
\begin{equation}
    \frac{\partial P(v,t)}{\partial t}=\gamma\frac{\partial}{\partial v} (vP(v,t)) +
    \frac{k_BT\gamma}{m}\frac{\partial^2P(v,t)}{\partial v^2} \,,
    \label{eq:FP}
\end{equation}
where $m$, $\gamma$, $T$ and $k_B$ are the particle's mass,
the friction coefficient, the temperature of the system and the Boltzmann constant, respectively.
Equation \eqref{eq:FP} was independently derived by 
Fokker \cite{fokker1913over,fokker1914mittlere}, Planck \cite{planck1917satz}, and Smoluchowski \cite{smoluchowski1913einige}, 
building on the earlier insight of Lord Rayleigh, who investigated the velocity dynamics of a massive particle $m$
colliding with a bath of lighter particles in thermal equilibrium at temperature $T$. This setting, later
known as the Rayleigh piston \cite{hoare1971linear,driessler1981spectrum}, leads at the macroscopic level to linear damping, i.e.,
$\dot v=-\gamma v$ \cite{stokes1851effect}, 
and, upon the introduction by Langevin 
of a stochastic force $\xi(t)$, to the
classical description of Brownian motion \cite{langevin1908comptes}
\begin{subequations}
\begin{align}
\dot{x}(t) &= v(t)
    \label{eq:kinetic} \\
\dot{v}(t) &= -\gamma \, v(t) + \xi(t) \,.
    \label{eq:langevin}
\end{align}
\end{subequations}
The statistical properties of the velocity $v(t)$ are entirely determined by
those of the stochastic force $\xi(t)$. In most physical applications, 
one specifies only $\langle\xi(t)\rangle=0$ and 
$\langle\xi(t_1)\xi(t_2)\rangle=\frac{2\gamma k_B T}{m} \,\delta(t_1-t_2)$, 
thereby ensuring consistency with the fluctuation--dissipation theorem
\cite{kubo1966}.
However, full
equivalence between the Langevin description~\eqref{eq:langevin} and the
associated Fokker--Planck equation~\eqref{eq:FP} requires the process $v(t)$
be Gaussian at all times. This, in turn, demands that $\xi(t)$ be Gaussian white
noise, implying an additional postulate beyond the first two moments
\cite{van1976stochastic} (see the Supplementary Material -- SM).

As a Gaussian process, all statistical properties of $v(t)$ are completely
determined by its two-time correlation function. 
In particular, any $2n$-time 
velocity correlation admits a Wick decomposition \cite{wick1950evaluation},
\begin{equation}
\langle v(t_1)\cdots v(t_{2n})\rangle
= \sum_{\text{all pairings}}
\prod_{\substack{1\le i<j\le 2n \\ (i,j)\,\text{paired}}}
\langle v(t_i)v(t_j)\rangle \,,
\label{eq:Wick_symb}
\end{equation}
with $(2n-1)!!$ distinct pairings. 
In the long-time limit, 
or under thermal initial conditions, 
velocity $v(t)$ 
reduces to the unique stationary Gaussian Markov
process with exponential autocorrelation, i.e.,
\begin{equation}
    \langle v(t_1)v(t_2)\rangle=\frac{k_BT}{m} \rm{e}^{-\gamma\left|t_1-t_2\right|} \,,
    \label{eq:OU_vel_CF}
\end{equation}
that is to the Ornstein--Uhlenbeck process
\cite{uhlenbeck1930theory,doob1942brownian,doob1944elementary}.
Therefore, also the particle's trajectory $x(t)$, 
determined by the linear kinematic relation
\eqref{eq:kinetic}, is Gaussian. Its even moments follow
directly from Wick’s theorem and obey the identity
$\langle x^{2n}(t)\rangle=(2n-1)!!\,\langle x^2(t)\rangle^n$. 
However, unlike
$v(t)$, $x(t)$ exhibits non-Markovian dynamics due to the 
time-correlation of the velocity \eqref{eq:OU_vel_CF}
\footnote{In general, time integration of a Markov process generates 
non‑Markovian dynamics, except in the singular case of white noise. 
We recall that Markovianity, as well as stationarity or ergodicity, 
is a property of a specific observable, 
not an intrinsic property of the underlying stochastic process.}.
Thus, the coupled system defined by Eqs.~\eqref{eq:kinetic} and \eqref{eq:OU_vel_CF}
provides a prototypical example of a non-Markovian Gaussian process. 
At large elapsed times, i.e., $t\gg 1/\gamma$, 
or in the large damping limit, 
the probability density function of the particle's position $P(x,t)$ 
obeys the standard Smoluchowski diffusion equation
\cite{smoluchowski1913einige,van1976stochastic,risken1989fokker}. 
At elapsed times smaller than the time-scale $1/\gamma$, 
the correct evolution equation still remains unavailable.
In this Letter we derive the \emph{exact diffusion equation} governing $P(x,t)$ for an
arbitrary Gaussian non-Markovian velocity process, not necessarily stationary.
Rather than restricting to Ornstein--Uhlenbeck statistics 
\eqref{eq:OU_vel_CF}, we consider a general
Gaussian velocity process $v(t)$ fully characterized by its two-time
correlation function $\langle v(t_1)v(t_2)\rangle$, thereby encompassing a wide
class of internal and external colored noises
\cite{van1976stochastic,mori1965transport,haunggi1994colored}.
Although linear systems driven by Gaussian colored noise have been extensively
studied
\cite{haunggi1994colored,van1976stochastic,zwanzig2001nonequilibrium,risken1989fokker},
our starting point follows Balescu’s \emph{physicist’s compromise}
\cite{balescu1997statistical,balescu2007v,dieterich2015fluctuation}, whereby the
velocity $v(t)$ is treated as a prescribed Gaussian process, rather than being
generated explicitly by an underlying Langevin equation, i.e., Eq. \eqref{eq:langevin}.

Within the Balescu's hybrid kinetic framework, 
the quasi-linear approximation leads to
the non-Markovian diffusion equation
\begin{equation}
    \frac{\partial P(x,t)}{\partial t}
    =
    \int_0^t dt_1 \,
    \left\{ \langle v(t)v(t_1)\rangle
    \frac{\partial^2 P(x,t_1)}{\partial x^2} \right\} \,,
    \label{eq:ZB}
\end{equation}
which first emerged from projection-operator coarse-graining of Hamiltonian
systems by Zwanzig
\cite{zwanzig1961memory,nordholm1975systematic,zwanzig2001nonequilibrium}.
We therefore refer to Eq.~\eqref{eq:ZB} as 
the \emph{Zwanzig--Balescu equation} (ZBE).

An alternative route, 
based on a non-Markovian generalization of the Chapman--Kolmogorov equation, 
was developed by H\"anggi 
\cite{hanggi1978correlation,hanggi1985functional,
hanggi1989colored,haunggi1994colored}
and leads to
\begin{equation}
    \frac{\partial P(x,t)}{\partial t}
    =
    \left\{ \int_{t_0}^{t} dt_1 \,
    \langle v(t)v(t_1)\rangle \right\}
    \frac{\partial^2 P(x,t)}{\partial x^2} \,.
    \label{eq:BH}
\end{equation}
This equation was originally introduced by Batchelor
in the context of turbulent diffusion \cite{batchelor1952diffusion}.
We therefore refer to Eq.~\eqref{eq:BH} as the
\emph{Batchelor--H\"anggi equation} (BHE).

In both equations, $v(t)$ is assumed to be Gaussian and stationary. However, the BHE suffers from well-known shortcomings: it depends explicitly on the initial time $t_0$ and may violate positivity \cite{fox1977generalized,haunggi1994colored}.
 More fundamentally, \emph{neither equation is exact}: both fail to reproduce Gaussian position moments beyond the second order, as shown in the SM. In particular, the ZBE does not yield the correct Gaussian kurtosis (i.e., equal to 3), while the BHE describes a Gaussian position process if and only if $\langle v(t)v(t')\rangle = D(t)\delta(t-t')$, i.e., in the Markovian limit.
 These structural shortcomings motivate the exact diffusion equation derived in the next section.


\emph{The truncated characteristic function hierarchy.}
We introduce the characteristic function of the
particle's distribution as 
$\widehat{P}(q,t)=\int_{-\infty}^{\infty} \! dx \, P(x,t)\, e^{iqx}$    
whose Taylor expansion around $q=0$ reads
\begin{equation}
    \widehat{P}(q,t)
    = \sum_{n=0}^{\infty} \frac{(iq)^{2n}}{(2n)!}\,
    \langle x^{2n}(t)\rangle \,.
    \label{eq:char_func_taylor}
\end{equation}
Taking the time derivative of Eq.~\eqref{eq:char_func_taylor} yields the exact identity
\begin{align}
    \frac{\partial}{\partial t}\widehat{P}(q,t)
    =
    \sum_{n=1}^{\infty}
    (iq)^{2n}\int_0^t dt_1 \int_0^{t_1} dt_2 \cdots\nonumber\\
    \int_0^{t_{2n-2}} dt_{2n-1}
    \langle v(t)v(t_1)\cdots v(t_{2n-1})\rangle \,,
    \label{eq:char_func_der}
\end{align}
with $t_0\equiv t$ (see SM). Eq.\eqref{eq:char_func_der} holds for any stochastic process, independently of any assumption of 
Markovianity, Gaussianity, or stationarity 
for the trajectory $x(t)$ or its velocity $v(t)$
(with the exception of Lévy–stable processes).
We now define the \emph{moment--truncated characteristic function}
(which coincides with the truncated cumulant expansion in the present Gaussian
setting) as
\begin{equation}
    \widehat{\rho}_N(q,t)
    =
    \sum_{n=0}^{N}
    \frac{(iq)^{2n}}{(2n)!}
    \langle x^{2n}(t)\rangle \,,
    \label{eq:trunc_char_func_taylor}
\end{equation}
with the consistency condition
$ \lim_{N\to\infty} \widehat{\rho}_N(q,t)
    = \widehat{P}(q,t)$.
Finally, by making use of the Wick decomposition \eqref{eq:Wick_symb}, the time evolution of
the truncated characteristic function can be written as
\begin{align}
    \frac{\partial}{\partial t}\widehat{\rho}_N(q,t)
    =&
    \sum_{n=1}^{N}
    (iq)^{2n}
    \sum_{\text{all pairings}}
    \int_0^t dt_1 \int_0^{t_1} dt_2 \cdots
    \nonumber\\
    &
    \int_0^{t_{2n-2}} dt_{2n-1}
    \prod_{\substack{0\le i<j\le 2n-1 \\ (i,j)\,\text{paired}}}
    \langle v(t_i) v(t_j) \rangle \,.
    \label{eq:trunc_char_func_der_gauss}
\end{align}
This equation admits a natural diagrammatic interpretation in terms of Wick contractions: each term in the sum \eqref{eq:Wick_symb} corresponds to a specific pairing of velocity variables and can be represented by a diagram in which lines connect the associated times. In this representation, the full $2n$‑time correlation function is given by the sum over all possible pairwise connections among the points $t_1,t_2,\ldots,t_{2n}$, each link denoting a two‑time correlation 
$\langle v(t_i)v(t_j)\rangle$. 
By instance, the $4$-times velocity correlation function is depicted as

\begin{equation}
\begin{aligned}
&\langle v(t_1)v(t_2)v(t_3)v(t_4)\rangle
=
\underbrace{%
\begin{tikzpicture}[baseline=-0.4ex]
\node (1) at (0,0) {$t_1$};
\node (2) at (1.2,0) {$t_2$};
\node (3) at (2.4,0) {$t_3$};
\node (4) at (3.6,0) {$t_4$};
\draw (1) to[bend left=45] (2);
\draw (3) to[bend left=45] (4);
\end{tikzpicture}}_{\;\langle v(t_1)v(t_2)\rangle\langle v(t_3)v(t_4)\rangle\;}
\;+\;\\
&\underbrace{%
\begin{tikzpicture}[baseline=-0.4ex]
\node (1) at (0,0) {$t_1$};
\node (2) at (1.2,0) {$t_2$};
\node (3) at (2.4,0) {$t_3$};
\node (4) at (3.6,0) {$t_4$};
\draw (1) to[bend left=45] (3);
\draw (2) to[bend left=45] (4);
\end{tikzpicture}}_{\;\langle v(t_1)v(t_3)\rangle\langle v(t_2)v(t_4)\rangle\;}
\;+\;
\underbrace{%
\begin{tikzpicture}[baseline=-0.4ex]
\node (1) at (0,0) {$t_1$};
\node (2) at (1.2,0) {$t_2$};
\node (3) at (2.4,0) {$t_3$};
\node (4) at (3.6,0) {$t_4$};
\draw (1) to[bend left=45] (4);
\draw (2) to[bend left=45] (3);
\end{tikzpicture}}_{\;\langle v(t_2)v(t_3)\rangle\langle v(t_1)v(t_4)\rangle\;}
\end{aligned}
\end{equation}

\begin{figure*}[t]
    \centering
    \includegraphics[width=\textwidth]{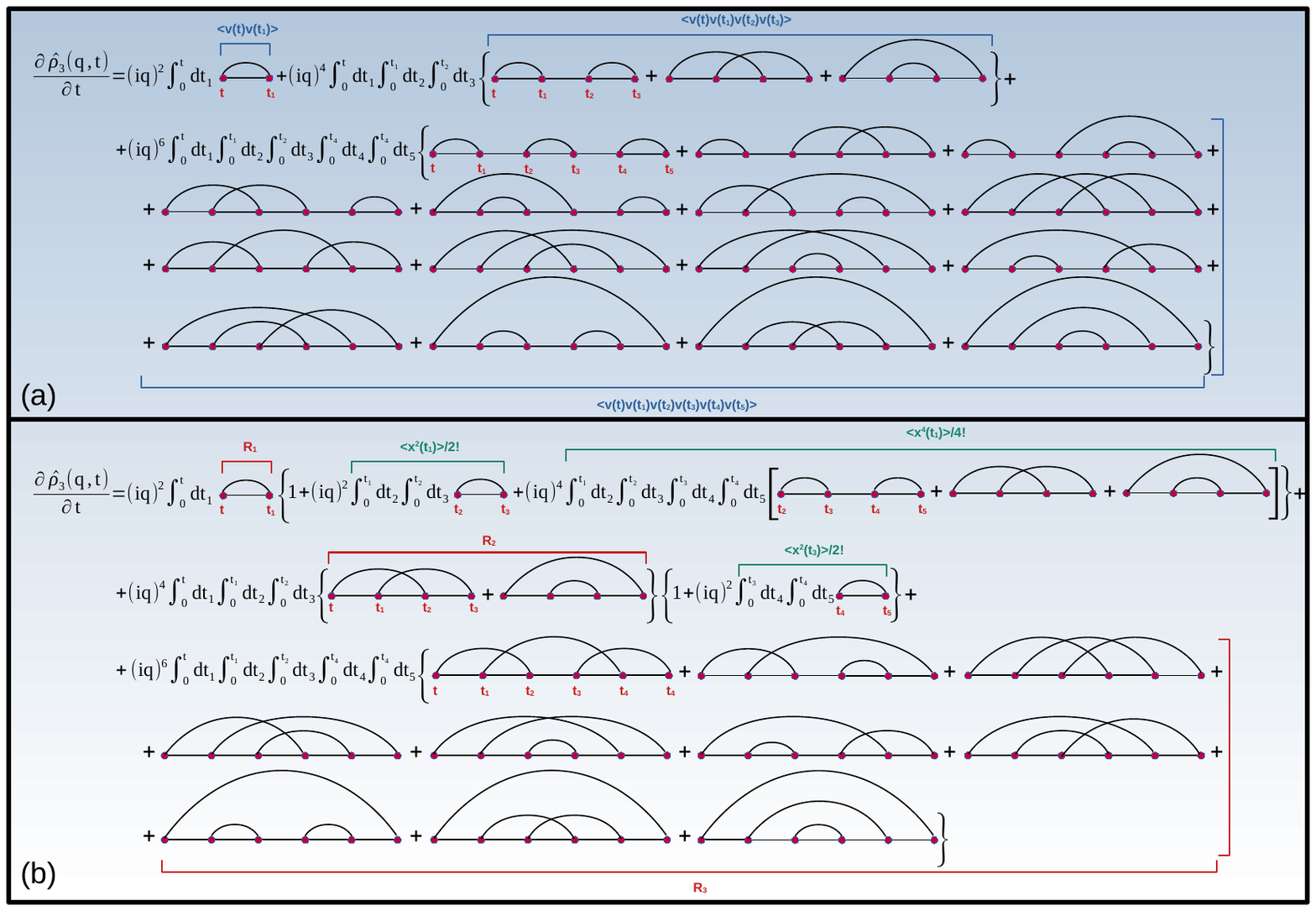}
    \caption{\emph{Diagrammatic representation of the moment--truncated characteristic function.}
(a) Equation~\eqref{eq:trunc_char_func_der_gauss_wick} admits a graphical representation
in terms of Wick pairings, as encoded by the contraction rule~\eqref{eq:Wick_symb}.
Each diagram corresponds to a distinct pairing of the time indices
$\{t_i,t_j\}$ appearing in the Gaussian decomposition. For truncation order $N=3$, the time evolution of the characteristic function is given
by the sum of $1$, $3$, and $15$ distinct diagrams, corresponding respectively to the
possible pairings of $2$, $4$, and $6$ time variables. Each diagram represents a product
of two-time velocity correlation functions $\langle v(t_i)v(t_j)\rangle$, 
associated with one complete pairing. 
(b) The same diagrammatic structure introduced in panel (a) is rearranged in order to
recover the compact expression~\eqref{eq:trunc_char_func_der_gauss_wick}. 
The coefficients $R_k$, encode the connected Wick contractions of order $2k$. 
They multiply the hierarchy of moment--truncated characteristic functions of
decreasing order, and are obtained as sums over geometrically connected Wick diagrams involving
$2k$ time arguments, in direct analogy with the linked‑cluster structure of connected Feynman diagrams \cite{goldstone1957derivation,peskin2018introduction,kleinert2009path}. The  number of Wick's contractions contributing to the coefficients $R_k$ are: $\bar R_1=1$, $\bar R_2=3$, $\bar R_3=10$ according to the recurrence relation \eqref{eq:recurrence_connected_wick}.}
    \label{fig:fig_wick}
\end{figure*}
Fig.\ref{fig:fig_wick}(a) shows the Wick diagrammatic representation of 
Eq. \eqref{eq:trunc_char_func_der_gauss} 
for $N=3$. Each term of the sum is composed of $(2n-1)!!$ Wick's pairings, with $n\in[1,3]$. 
Now, It is possible to rearrange the diagrammatic structure of the sum on the RHS of Eq.\eqref{eq:trunc_char_func_der_gauss} regrouping the diagrams in such a way that 
\begin{align}
    &\frac{\partial}{\partial t}
    \widehat{\rho}_N(q,t)=\sum_{k=1}^{N} (iq)^{2k}\int_0^t dt_1 \int_0^{t_1} dt_2 \cdots\nonumber
    \\
    & \int_0^{t_{2k-2}} dt_{2k-1} R(t,t_1,\ldots,t_{2k-1})
    \sum_{j=0}^{N-k}
    \frac{(iq)^{2j}}{(2j)!} \langle x^{2j}(t_{2k-1})\rangle
    \,.
    \label{eq:trunc_char_func_der_gauss_wick}
\end{align}
This is illustrated in Fig.~\ref{fig:fig_wick}(b). 
The quantity $R(t,t_1,\ldots,t_{2n-1})$ denotes the sum of all geometrically
connected Wick diagrams on the ordered set $(t,t_1,\ldots,t_{2n-1})$. 
A Wick diagram is said to be connected if the time points cannot be split into
two non‑empty consecutive blocks such that no contraction links the two sets
(see the graphical definition in the SM). 
This notion follows directly the linked‑cluster construction of quantum field
theory and refers here solely to the geometric (topological) connectedness of
Wick contraction graphs
\cite{bogoliubov1959introduction,mayer1940statistical,kleinert2009path}.
In quantum field theory, 
only connected Feynman diagrams contribute to physical correlations, 
since disconnected diagrams cancel in normalized generating functionals. 
Consequently, 
the logarithm of the generating functional generates connected Green’s functions only, 
which determine observable quantities and ensure the linked‑cluster structure 
and consistency of the theory
\cite{peskin2018introduction,zinnjustin2002,itzykson1980,goldstone1957derivation}.

Let us adopt the following shorthand notation for 
the sum of connected Wick's diagrams of order $2n$: $R_n=R(t,t_1,\ldots,t_{2n-1})$. 
Then, thanks to \eqref{eq:trunc_char_func_taylor}, 
Eq.\eqref{eq:trunc_char_func_der_gauss_wick} becomes 
  \begin{align}
    \frac{\partial}{\partial t}\widehat{\rho}_N(q,t)
    &=\sum_{k=1}^{N} (iq)^{2k} \int_0^t dt_1 \int_0^{t_1} dt_2 \cdots
    \nonumber
    \\
    & \cdots\int_0^{t_{2k-2}} dt_{2k-1} R_k \, \widehat{\rho}_{N-k}(q,t_{2k-1}) \,.
\label{eq:trunc_char_func_der_gauss_wick_1}
\end{align}
This hierarchical recurrence relation among 
truncated characteristic functions constitutes the main result of our analysis. 

\emph{Combinatorial properties}. 
Equation~\eqref{eq:trunc_char_func_der_gauss} shows that 
the coefficient of the term $(iq)^{2n}$ is given by the total number of 
the associated Wick pairings: $(2n-1)!!$. 
As illustrated in Fig.~\ref{fig:fig_wick}, these pairings can be reorganized into the distinct contributions associated with the $R_k$ terms entering in
Eq.~\eqref{eq:trunc_char_func_der_gauss_wick}.
To formalize this regrouping, 
let $\bar R_k$ denote the number of connected Wick diagrams whose sum yields the term $R_k$. 
Matching the RHSs of Eqs.~\eqref{eq:trunc_char_func_der_gauss} and \eqref{eq:trunc_char_func_der_gauss_wick} requires that, 
for each power of $q$, the multiplicative diagrammatic must coincide. 
This condition selects only the contributions satisfying $n = k + j$ in Eq.~\eqref{eq:trunc_char_func_der_gauss_wick}, 
implying that the number of Wick pairings of order $2n$ satisfies the recursive relation
\begin{equation}
    (2n-1)!! =   \sum_{k=1}^{n-1} [\,2(n-k)-1\,]!!\, \bar R_k + \bar R_n \,,
    \label{eq:recurrence_connected_wick}
\end{equation}
with $\bar R_1 = 1$, since $R_1 \equiv \langle v(t)v(t_1) \rangle$. This rule precisely governs how the diagrams in panel (a) of Fig.~\ref{fig:fig_wick} are rearranged into the connected contributions shown in panel (b). By explicitly expanding the sum over $k$ in Eq.~\eqref{eq:recurrence_connected_wick}, one obtains the triangular structure summarized in the following Table 

\begin{table}[!ht]
\centering
\begin{tabular}{|l|llllll|}
  \hline
  $n \backslash k$ & 1 & 2 & 3 & 4 & 5 & 6 \\ \hline
  1 & $\bar R_1$ & ~ & ~ & ~ & ~ & ~ \\ 
  2 & $1!!\, \bar R_1$ & $\bar R_2$ & ~ & ~ & ~ & ~ \\ 
  3 & $3!!\,\bar R_1$ & $1!!\,\bar R_2$ & $\bar R_3$ & ~ & ~ & ~ \\ 
  4 & $5!!\,\bar R_1$ & $3!!\,\bar R_2$ & $1!!\,\bar R_3$ & $\bar R_4$ & ~ & ~ \\ 
  5 & $7!!\,\bar R_1$ & $5!!\,\bar R_2$ & $3!!\,\bar R_3$ & $1!!\,\bar R_4$ & $\bar R_5$ & ~ \\ 
  6 & $9!!\,\bar R_1$ & $7!!\,\bar R_2$ & $5!!\,\bar R_3$ & $3!!\,\bar R_4$ & $1!!\,\bar R_5$ & $\bar R_6$ \\ \hline
\end{tabular}
\label{tab:table_wick}
\end{table}
A simple rearrangement of Eq.~\eqref{eq:recurrence_connected_wick} 
yields the equivalent form 
$\bar R_n = (2n-1)!! - \sum_{k=1}^{n-1} (2k-1)!! \, \bar R_{\,n-k}$,
which coincides with the well-known recurrence relation for the number of connected Feynman diagrams of order $2n$ contributing to the QED electron propagator (two‑point function)~\cite{OEIS_A000698}. More broadly, it constitutes one of the classical identities satisfied by odd double factorials~\cite{callan2009combinatorial}.

\emph{The diffusion equation for non-Markovian Gaussian processes}. 
In the infinite truncation limit, the evolution equation
for the characteristic function is recovered in the form
\begin{align}
\frac{\partial}{\partial t}\widehat{P}(q,t)
&=
\sum_{k=1}^{\infty} (iq)^{2k}
\int_{0}^{t} dt_{1}
\int_{0}^{t_{1}} dt_{2}\cdots
\nonumber\\
&\qquad\cdots
\int_{0}^{t_{2k-2}} dt_{2k-1} \,
R_{k} \,
\widehat{P}(q,t_{2k-1}) \,.
\label{eq:char_func_gauss_full}
\end{align}
By inverting the Fourier transform, we finally obtain the exact evolution
equation for the probability density function of a non‑Markovian Gaussian
process:
\begin{align}
\frac{\partial}{\partial t} P(x,t)
&=
\sum_{k=1}^{\infty}
\int_{0}^{t} dt_{1}
\int_{0}^{t_{1}} dt_{2}\cdots
\nonumber\\
&\qquad\cdots
\int_{0}^{t_{2k-2}} dt_{2k-1}\,
R_{k}\,
\frac{\partial^{2k}}{\partial x^{2k}}
P(x,t_{2k-1}) \,.
\label{eq:pdf_gauss_full}
\end{align}
The first term of the summation, i.e., $k=1$, coincides with the ZBE, the second term yields the kurtosis equal to 3, but, in general, any finite truncation
destroys the Gaussian character of the solution.
Only the full infinite series restores Gaussianity exactly.
Remarkably, no assumption has been made on the stationarity of the velocity
process. Therefore, Eqs.~\eqref{eq:trunc_char_func_der_gauss} and
\eqref{eq:pdf_gauss_full} are valid for both stationary and non‑stationary
kinetics defined by Eq.~\eqref{eq:kinetic}.

\emph{Ornstein--Uhlenbeck process.}
We now briefly discuss the specific case of the OU process defined by 
Eq.~\eqref{eq:OU_vel_CF}. The only equation that fully preserves the Gaussian
structure of $P(x,t)$ is Eq.~\eqref{eq:pdf_gauss_full}, 
which remains exact for arbitrary values of the
damping coefficient $\gamma$ and for all times, including the inertial
non-stationary transient. In the long-time or large-damping limit, one recovers
the Smoluchowski diffusion equation, corresponding to
$R_1 \to D\,\delta(t-t_1)$ while all higher-order contributions
($k\geq 2$) vanish (see SM).
Starting from the Klein--Kramers equation~\cite{klein1921statistischen}, 
a systematic derivation of the Smoluchowski equation, together with its
higher-order corrections in powers of $\gamma^{-1}$, is presented in the book of Risken~\cite{risken1989fokker}. 
The leading term of the Risken expansion coincides with the (ZBE), 
whereas the
higher-order corrections differ significantly from the coefficients appearing
in Eq.~\eqref{eq:pdf_gauss_full}.

\emph{Ornstein--Uhlenbeck active noise.}
When the noise in Eq.~\eqref{eq:langevin} is exponentially correlated, one speaks of an Ornstein--Uhlenbeck noise, i.e. $\langle \xi(t_1)\xi(t_2)\rangle= v_p^2\gamma^2e^{-\frac{|t_1-t_2|}{\tau}}$. In the long-time limit one has 

\begin{equation}
    \langle v(t_1)v(t_2)\rangle=\frac{v_p^2\tau^2\gamma^2}{1-\tau^2\gamma^2}\left[\frac{e^{-\gamma|t_1-t_2|}}{\gamma\tau}-e^{-\frac{|t_1-t_2|}
    {\tau}}\right]
    \label{eq:OU_noise}
\end{equation}

\noindent (see the SM for the full expression including transients). This model has been widely used over the past decade to capture the coarse-grained motion of an active particle, where $v_p$ is the propulsion speed and $\tau$ is the persistence (correlation) time \cite{bonilla2019active,maggi2015multidimensional,martin2021statistical,fodor2018statistical}. The stochastic term $\xi(t)$ is a Gaussian colored noise representing an athermal energy input (external noise), and therefore does not satisfy the fluctuation--dissipation relation. This entails a non-Markovian Gaussian dynamics for both the velocity $v(t)$ and the position $x(t)$ \cite{baek2023markovian}. As a consequence, Eq.~\eqref{eq:pdf_gauss_full} provides the exact evolution equation for the position pdf of an active Ornstein--Uhlenbeck particle.

\emph{Generalized Langevin equation.} A natural generalization of the standard Langevin dynamics is provided by the generalized Langevin equation \cite{mori1965transport,zwanzig2001nonequilibrium,van1976stochastic}, which incorporates memory effects through a non-local friction kernel:
\begin{equation}
\dot{v}(t) = -\int_0^t ds\, \gamma(t-s)\, v(s) + \xi(t).
\end{equation}
Here $\gamma(t)$ is a memory kernel and $\xi(t)$ is a Gaussian stochastic process. In equilibrium, the fluctuation--dissipation theorem imposes $\langle \xi(t_1)\xi(t_2) \rangle = k_B T\, \gamma(|t_1-t_2|)$. The resulting dynamics is Gaussian and non-Markovian, with velocity autocorrelation $\tilde C_v(s)=\frac{k_B T/m}{s+\tilde\gamma(s)}$ in Laplace space. Also in this case, Eq.~\eqref{eq:pdf_gauss_full} provides the exact evolution equation for $P(x,t)$.

\emph{Fractional Brownian motion.}
Fractional Brownian motion is a Gaussian process characterized by long--range,
stationary velocity correlations of the form $
\langle v(t_1)v(t_2)\rangle
=
2\chi H(2H-1)\,|t_1-t_2|^{2H-2}$,
where $0<H<1$ is the Hurst exponent and $\chi>0$
\cite{mandelbrot1968fractional}.
Contrary to the conclusions of Ref.~\cite{wang1990long}, 
which coincides with the BHE, 
Eq.~\eqref{eq:pdf_gauss_full} provides the correct evolution
equation for the displacement probability density $P(x,t)$.
In the persistent regime, $1/2<H<1$, the first term of the sum
\eqref{eq:pdf_gauss_full} can be derived explicitly, as shown in the SM, yielding the fractional wave-equation 
\cite{schneider_etal-jmp-1989,mainardi-csf-1996}
\[
\mathcal{D}^{2H} P 
\simeq
\chi \, \Gamma(2H+1) \,
\frac{\partial^2 P}{\partial x^2} \,,
\quad 1 < 2H < 2 \,,
\]
where $\mathcal{D}^\mu$, with $\mu>0$, denotes the 
time-fractional derivative of order $\mu$
in the Caputo sense \cite{podlubny1998fractional}.
Since $\mathcal{D}^1\equiv d/dt$, the standard diffusion equation with diffusion coefficient $\chi$ is recovered when $H=1/2$.
In the antipersistent regime, $0<H<1/2$, the velocity correlation kernel becomes
non--integrable at coinciding times, and the derivation of an analogous first--order
equation remains an open problem that we will address in a forthcoming paper.

\begin{acknowledgments}
AT thanks prof. D. Callan for having pointed out the existing relation analogous to \eqref{eq:recurrence_connected_wick} and acknowledges illuminating discussions with prof. F. Marchesoni.
\end{acknowledgments}

\bibliographystyle{apsrev4-2}
\bibliography{diffusion}

\end{document}